\documentclass[%
 reprint,
 aip,
 jmp,
 amsmath,amssymb,
]{revtex4-2}

\usepackage{graphicx}
\usepackage{dcolumn}
\usepackage{bm}
\usepackage[utf8]{inputenc}
\usepackage[english]{babel}

\begin{document}

\preprint{APS/123-QED}

\title{Continuous slowing of a gadolinium atomic beam}

\author{A. Chavarr\'{i}a--Sibaja}
\email{andres.chavarriasiabaja@ucr.ac.cr}

\author{ A. Araya--Olmedo}
\email{andre.araya@ucr.ac.cr}
\affiliation{Escuela de F\'isica, Universidad de Costa Rica, 2060 San Pedro, San Jos\'e, Costa Rica.}%
\affiliation{Centro de Investigaci\'on en Ciencia e Ingenier\'ia de Materiales, Universidad de Costa Rica, 2060 San Pedro, San Jos\'e, Costa Rica}%

\author{O.A. Herrera--Sancho}
\email{oscar.herrerasancho@ucr.ac.cr}

\affiliation{Escuela de F\'isica, Universidad de Costa Rica, 2060 San Pedro, San Jos\'e, Costa Rica.}%
\affiliation{Centro de Investigaci\'on en Ciencia e Ingenier\'ia de Materiales, Universidad de Costa Rica, 2060 San Pedro, San Jos\'e, Costa Rica}%
\affiliation{Centro de Investigaci\'on en Ciencia atómicas, nucleares y moleculares, Universidad de Costa Rica, 2060 San Pedro, San Jos\'e, Costa Rica}%

\email{oscar.herrerasancho@ucr.ac.cr}

\date{\today}

\begin{abstract}

The article presents the development of a new and innovative experimental method to fully characterize a solenoidal ''spin-flip" Zeeman slower (ZS) using a Quartz Crystal $\mu$-balance (QCM) as a kinetic energy sensor. In this experiment, we focus a 447.1 nm laser into a counter-propagating beam of gadolinium (Gd) atoms in order to drive the dipole transition between ground $^{9}$D$^{0}_{2}$ state and $^{9}$D$_{3}$ excited state. The changes in the velocity of the beam were measured using a QCM during this process, as a novel and alternative method to characterize the efficiency of a 1 m-long spin-flip Zeeman slower. The QCM, normally used in solid-state physics, is continuously and carefully monitored to determine the change in its natural frequency of oscillation. These changes reveal a direct relation with changes in the deposition rate and the momentum exchanged between the QCM and Gd atoms. Hence, in terms of ultracold atom physics, it might be used to study the time-evolution of the velocity distribution of the atoms during the cooling process. By this method, we obtain a maximum atom average velocity reduction of (43.5 $\pm$ 6.4)$\%$ produced by our apparatus. Moreover, we estimate an experimental lifetime of $\tau_{e}$ = 8.2 ns for the used electronic transition, and then we compared it with the reported lifetime for 443.06 nm and 451.96 nm electronic transitions of Gd. These results confirm that the QCM offers an accessible and simple solution to take into account for laser cooling experiments. Therefore, a novel and innovative technique can be available for future experiments.   

\end{abstract}


\maketitle

\section{Introduction}
\label{sec:Introduction}

 An important feat in atomic physics has been the diminishing of the kinetic energy (KE), or velocity of the atoms, by the production of innovations in experimental and fundamental theoretical fields. These advances have allowed the investigation of nearly ideal quantum systems such as cold plasmas and Bose-Einstein condensates of dilute gases in ultracold atom physics~\cite{wieman}. Techniques like Doppler cooling, Chirp cooling, Magneto-Optical Traps (MOT), and others~\cite{foot,veit2020pulsed_microscope,Guggemos_2019,PhysRevA.88.012512} have been used to control the position and velocity of the atoms in order to obtain such systems. The atomic manipulation of these methods is based, among other principles, on momentum exchange between atoms and the photons during an electronic transition in a magnetic field. However, different atomic properties like frequency detuning generated by Doppler shift, sensibility to hyperfine structures~\cite{foot}, and the forbidden transition~\cite{landini2011, rosi} causes its study to be in constant development in the physics of the ultracold atoms. As consequense, the advancements by the three 1997 Nobel prize winners S. Chu~\cite{chu}, C. Cohen-Tannoudji~\cite{cohen}, and W.D. Phillips~\cite{phillips} that results in the development of Zeeman slower (ZS), have been undeniably valuable to solve the detuning limitations for slowing atoms~\cite{davis_etal}. The function of the ZS is to maintain the resonance between the cooling laser frequency and the frequency of a selected electronic transition of an specific atom. Then, the cooling laser can interact with incoming evaporated particles, recoiling them and, therefore, reducing its velocity with minimum detuning. The apparatus maintains the resonance correcting the Doppler shift by a magnetic field that generates and adjusts the split of electronic energy levels of an atom (Zeeman effect) so that the frequency of the electronic transition is always close to that of the laser frequency~\cite{foot,1_chambers2004modern, bober, 2_guevara2016diseno,tempelaars}. 

The ZS has been used for the cooling of atoms like Rubidium~\cite{anderson1995observation}, and Sodium~\cite{davis1995bose}. Likewise, in the last years the lanthanides elements like Ytterbium (Yb)~\cite{xu2009laser,bell1991laser}, Erbium (Er)~\cite{berglund2007sub,mcclelland2006laser} and Dysprosium (Dy)~\cite{leefer2010transverse,lu2010trapping} have been explored for laser cooling experiments. These elements are explored due to its higher magnetic momentum ( $\mu$ $\leq$ 10 $\mu_{B}$)~\cite{leefer2010transverse, battles2020laser}, that allows an extensive investigation of the short-range interaction and the anisotropic dipolar interaction~\cite{baranov2012condensed,stuhler2005observation}. However, other lanthanides elements like gadolinium (Gd) remain an open research field for use in laser cooling. Recent experiments have explored laser cooling of Gd atoms and the measurement of spin forbidden transitions as a possible application towards new optical frequency standards in optical clocks~\cite{battles2020laser,upendra2019laser}. Moreover, due to its complex electronic configuration, Gd has properties that could be of interest for the study of atom-atom interaction of the strongly correlated matter~\cite{adhikari2019thesis,battles2020thesis}. Nevertheless, the novelty of the study of the use of Gd atoms in laser cooling makes this an open field for experimental and theoretical exploration.

The influence that these contributions have had in solid-state physics has resulted in a wide range of applications, from the study of surface science~\cite{guevara2016leed,lewenstein2007ultracold}, to many-body physics, including preparation and characterization of correlated phenomena, e.g. optical lattices~\cite{leblanc,guevara2016tesis, morsch, bloch}. Thus, the shared use of techniques from both fields of physics can make possible the creation of experimental tools that can allow the understanding of different phenomena typical of both fields of study. Additionally, this combination can be useful to create a new characterization method for the efficiency of the radiation-based cooling process. Therefore, an alternative method could be developed to detect the velocity change of the atoms in the cooling experiments by using tools from solid-state physics to measure atomic behaviour. The Quartz Crystal $\mu$-balance (QCM) is a simple tool to implement in experimental solid-state, normally used in processes of growing thin-films of diverse materials by evaporation. The principle of operation of the QCM is based on the continuous monitoring of its natural frequency of oscillation~\cite{bruckenstein1985experimental,boutamine2014hexamethyldisiloxane} due to the increase or decrease of the deposited mass on its surface. As mentioned in other research~\cite{3_guevara2018energy}, the oscillation frequency of the QCM could be affected by the momentum exchange of the atoms when they hit the surface of the crystal. This source of disturbance is indirectly related to the velocity of the deposited atoms, and it can be affected by the increase or decrease of the deposition rate of the atoms in the QCM. As a result, it is possible to translate the momentum exchange from the incoming atoms in percentage of increase or decrease of the velocity of the beam of atoms. Then, the verification of this deceleration could be confirmed by a computational model based on the Fokker-Planck equation (FPE). The FPE has mathematical properties that ease the study of dynamic variables and have been previously applied in atom cooling~\cite{6_nienhuis, 5_hoogerland}. The QCM implementation and the use of a numerical model achieves a different approach to the atom's velocity detection that dispenses the use of high-technology equipment to report these velocities. Thus, the development of the technique purpose in this article could represent an alternative that facilitates the creation of ultracold atom physics experiments in laboratories where, due to technical limitations, there might be a lack of detection equipment traditionally used to measure the velocities of a beam of atoms.

In this article, we present a novel and alternative method to characterize an efficient spin-flip ZS in an experiment with a beam of Gd atoms. This technique is presented as an alternative method for the measurement of the change of velocity of the beam of atoms in a cooling process that can be easily implemented by the use of a solid-state QCM. The development of the technique purpose in this article could represent an alternative that facilitates the creation of ultracold atom physics experiments in laboratories where, due to technical limitations, there might be a lack of detection equipment traditionally used to measure the velocities of a beam of atoms. Our results showed that the QCM has sufficient measurement resolution to estimate, at least by order of magnitude, the deceleration produced in atom cooling experiments and could be considered as a possible new mechanically based measurement technique for this type of experiment. As a short tour of the article, in section II, we present the numerical model used based on the FPE, and prioritize how the model translates to the experimental observations of atom deceleration. In section III, the experimental setup is described, and, in sections IV and V, we explain the experimental methods carried out during the trials and the analysis of the data obtained. In section VI, the results are discussed and compared with those of the numerical model and, in section VII, we set the conclusions and outlook of the experiment are set.

\section{Numerical Model}
\label{sec:Numerical_Model}

As mentioned before, we used a QCM to measure the variation of velocities. Therefore, a numerical model was implemented to simulate the deceleration of the atoms and verify the experimental results. This method is based on the Fokker-Planck equation (FPE), which is a stochastic partial differential equation that has been used to model the changes in the velocity distributions of the atom beam generated by the cooling process in different experiments~\cite{zoll}.

\begin{figure}
\begin{center}
\includegraphics[width=1.0\textwidth]{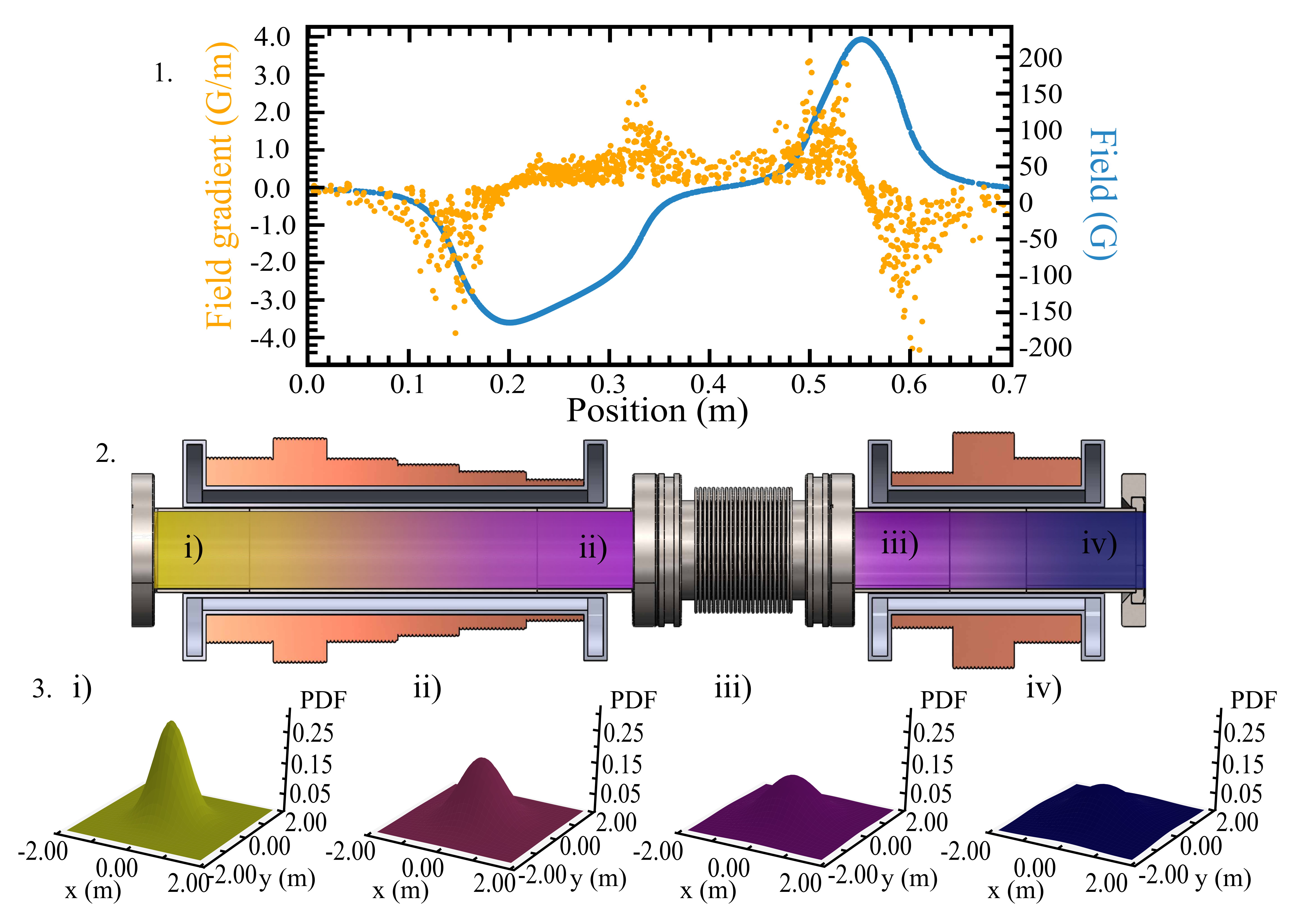}
\caption{Panel 1. Gradient for the magnetic field ($dB/dz$) and magnitude of the magnetic field generated by the ZS at 4 A (orange dots and blue dots, respectively). The showed gradient values were in a range between 4.39$\times$10$^{-4}$ G/m t and -3.31 $\times$ 10$^{-4}$ G/m, and corresponds to a magnetic field with values between -220 G to 220 G for a current of 4 A. Panel 2 and Panel 3. Time evolution of the velocity distribution of particles inside the ZS. The z-axis represents the normalized probability density function (PDF) of the atom velocity in arbitrary units. The x-axis and y-axis correspond to the axes of the cross-section of the ZS. i) The atoms enter our Zeeman Slower (ZS) with a Gaussian distribution given by the spread of the Gd evaporation. ii) During the passage through the first stage, the gaussian distribution gets flattened due to the action of the laser cooling. iii) The process continues after the spin flip section and the bell continues to flatten in a more pronounced way due to an increased magnetic field gradient. iv) The velocity distribution reaches its maximum flattening at the end of the ZS. The flattening of the center of the bell is an expression of a slowdown process of the atoms since the interactions happen on the center portion of the distribution. The particles in this region slow down, and consequently, the Gaussian surface flattens, causing the atoms to have a more uniform distribution due to their reduced velocity.}
\label{fig:fpe}
\end{center}
\end{figure}

The numerical model use a semiclassical approach of laser cooling to associate the FPE with the physical phenomenon of atom cooling. This approach based on a quantum mechanical treatment of the interaction between the light field with atoms. As is considered in other studies~\cite{4_smeets,5_hoogerland,6_nienhuis}, the model take into account a Brownian motion for the trajectory of the atoms. This approach is valid given that for Gd atoms, the velocities are not in the same order of magnitude as the recoil velocity, $v_r = \hbar k/M$, where $\hbar$ is the reduced Planck constant ($\hbar$ = $\frac{h}{2\pi}$), $k$ is the wavenumber, and $M$ is the atomic mass. Therefore, the velocity distribution, \textit{W(v)}, of the atoms is described by the FPE:

\begin{equation}
    \frac{\partial}{\partial t}W(v) = -\frac{1}{M}\frac{\partial}{\partial v}F(v)W(v) + \frac{1}{M^2}\frac{\partial ^2}{\partial v \partial v}D(v)W(v), \label{eq:fpe} \centering
\end{equation}

where \textit{F(v)} is the velocity dependant force, and \textit{D(v)} is a diffusion coefficient. Here, the first term of the equation is negative given that the force upon the incoming atoms produces a deceleration effect, while the diffusion coefficient adds variability to the spread of the incoming atoms. For the extent of this paper, the diffusion coefficient is taken as Gaussian noise given that the atoms' movement is considered as Brownian. This assumption is valid because the characteristics of the experimental setup generate a mean free path greater than the size of the space of interaction~\cite{4_smeets}. The velocity dependant force does not evolve directly with the time solution of the FPE. Rather, it is calculated by propagating the time solution through the experimental setup, so that the possible atom-photon interactions, the Zeeman effect, and the velocity itself are taken into account. These last phenomena are products of the Doppler effect given the interaction between the atoms and the photons, where the Zeeman effect is a compensation for the possible detuning of this event.

The matrix numerical method proposed and describes in previous investigations~\cite{holubec} was employed to solve the FPE in our numerical model. Here, the plane in which the atoms travel is discretized so that a transition rate is found given the probability of transition between adjacent points in the lattice. Then, the time evolution of the velocity distribution is obtained by a time-ordered exponential that sums over all the paths possible in the discretized space.

Note that the previous algorithm finds the evolution of the FPE as if it is stationary, e.g. not moving through the experimental setup. Therefore, the solution is propagated through the setup so that the velocity through any given point inside the space of the ZS can be determined. Thus, it is possible to obtain the final velocity of the atoms deposited in the QCM. Figure~\ref{fig:fpe} shows the results of the numerical model, where a flattening of the atomic velocity distribution occurs as the atoms are decelerated. Here, the subfigure 3.i) corresponds to the velocity distribution of the atoms at the entrance of the Zeeman Slower, 3.ii) and 3.iii) show the flattening of the same distribution before the spin-flip and after the spin-flip, respectively, and 3.iv) shows the final distributions when atoms are at the QCM. The flattening of the surface indicates the slowing of the atoms' by spreading the distribution through time.

\section{Experimental Setup}
\label{sec:Experimental_setup}

\begin{figure}
\begin{center}
\includegraphics[width=0.9\textwidth]{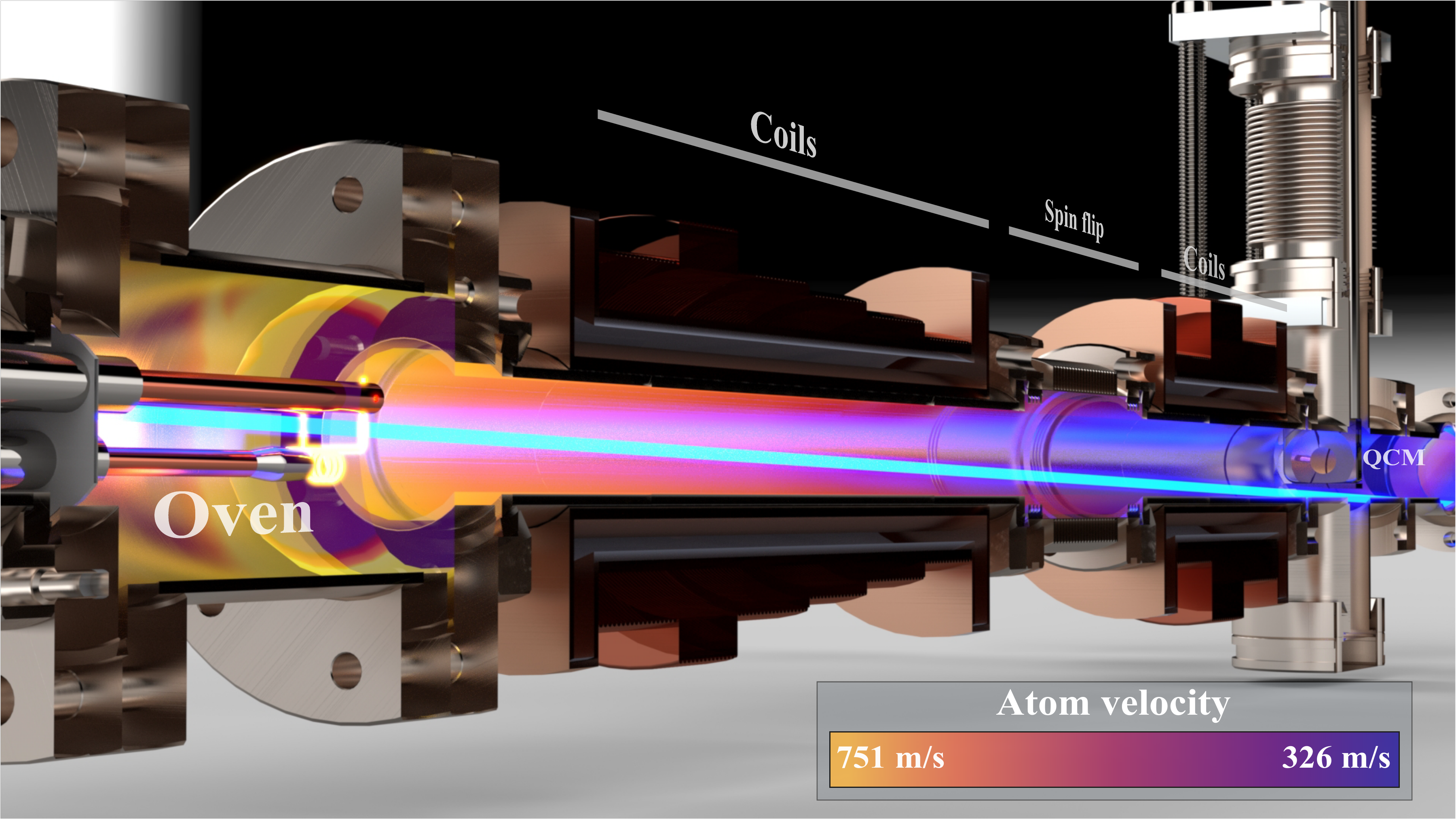}
\caption{Artistic visualization of the experimental setup used for the measurement of the slowdown effect of the Gd atoms using the variation of frequency of the QCM. The technical details of our ZS are exposed in previous research~\cite{2_guevara2016diseno}. The oven (left side of the figure) produces an atom beam using a hot filament. The velocity of the beam is reduced $\sim$43.4 $\%$ when the beam reaches the final section of our ZS at maximum efficiency. The process is represented by the colored gradient beam inside the apparatus. Finally, the QCM (right side of the figure) measures the change of the momentum exchanged produced due to the interaction of the atoms with the laser of 447.1 nm and the magnetic field.}
\label{fig:Zeeman_working}
\end{center}
\end{figure}

Figure~\ref{fig:Zeeman_working} shows the scheme of our experimental setup for the cooling of Gd atoms. The experiment was done in a high vacuum system with total internal volume $\sim$ 4$\times10^{-4}$ m$^{3}$. The main part of the volume is used for the application of the magnetic field, while the rest is used by the source of atoms and the QCM, both placed inside the apparatus. The internal base pressure is maintained at $10^{-9}$ Torr by one ion pump model Varian Vacion Plus 25 Triode of 20 L/s capacity. This internal presure is constantly monitored by a hot-cathode Bayard-Alpert gauge model MDC-432006 during all of the measurements reported below. 

For the experiment, it was used a sample of $\sim$3.9$\times$10$^{-3}$ kg of neutral metallic Gd with 99.9 $\%$ purity. The Gd has Z = 64, and its most abundant isotopes are five bosonic isotopes: $^{152}$Gd (0.2 $\%$), $^{154}$Gd (2.15 $\%$), $^{156}$Gd (20.47 $\%$), $^{158}$Gd (24.87 $\%$), and $^{160}$Gd (21.90 $\%$); and another two fermionic isotopes: $^{155}$Gd (14.73 $\%$) and $^{157}$Gd (15.68 $\%$)~\cite{sankari1999isotope_isotope_abundance, suryanarayana1997simulation?isotopic_abundance, eugster1970isotopic_Gd}. Also, Gd belongs to the Lanthanide group that has $4f$ and $5d$ inner electronic shells partially filled~\cite{von1989electronic_of_Gd}. This partially filled shell generates a rich spectrum of emission and absorption produced by the superimposed electron configuration. At the same time, this electronic configuration makes it difficult to resolve the individual lines of the different electronic states. The Gd has a mass m$_{Gd}$ = 157.25 u, a melting point of 1586 K, and a boiling point of 3539 K \cite{laing2009gadolinium_Tv_and_Tb}. The assumption is that, the sample of Gd used in our experiment is formed mainly by atoms of Gd$^{158}$ and Gd$^{160}$. According to other research~\cite{niki1989hyperfine,niki2005measurement}, the isotopic shift for Gd$^{158}$ and Gd$^{160}$ ranges from 0.86 to 1.18 GHz. Other studies reports~\cite{adhikari2019hyperfine, ahmad1979isotope} an isotopic shift for the Gd in the order of a few GHz.

To heat and evaporate the atoms, it was used a hot filament oven as a source of atoms. The oven consists of a helical filament of 3.5 mm radius formed by three turns of Tungsten wire of 0.5 mm diameter, situated 1.0 cm in front of a high voltage electrode with an alumina crucible containing a sample of Gd. Meanwhile, during the experiment, the temperature was optically measured with the use of a UV-Vis photo-spectrometer model Avantes AvaSpec-3648 connected to an optical fiber situated in the viewport outside the vacuum system. This photo-spectrometer allowed us to measure temperatures up to 4052 K with uncertainties between 13.5$\%$ and 18.5$\%$. Also, the internal pressure of the system increased to 10$^{-8}$ Torr when the oven was operating.

Besides, we used a spin-flip Zeeman slower (ZS) that generates a magnetic field values between 53.5 G and 374.8 G, that correspond to current values between 1 A and 7 A, respectively. The magnetic field gradients generated in this range of current are between -8.3$\times$10$^{-5}$ G/m and -6.62$\times$10$^{-4}$ G/m for the initial section the ZS and between 1.1$\times$10$^{-4}$ G/m and 8.8$\times$10$^{-5}$ G/m for the final section of the ZS after the spin-flip. The technical details of the design and construction of the ZS can be found in previous research~\cite{2_guevara2016diseno}. During the experiment, we measured the current values applied to the ZS coils due to the impossibility of a direct measure of the magnetic field inside the apparatus. Also, the range of currents used was determined during the development of the experiments in order to map the efficiency of our system. 

Moreover, in addition to our ZS, we used a blue laser model Laserglow Polaris-100 with a wavelegth of 447.1 nm. As shown in Fig.~\ref{fig:Zeeman_working}, the laser was alined slightly offset from the central axis of the ZS with an angle of 1.3°. This alignment was due to the presence of the QCM, as shown in the Fig.~\ref{fig:Zeeman_working}. Furthermore, the tilt angle of the laser prevents the QCM from blocking beam, so that the QCM, the oven, and the laser beam were in the same plane. The above-mentioned promotes a greater number of interactions between the laser and the atom beam when the magnetic field is on. In addition, the laser has a Gaussian distribution emission centered at 447.10 nm (6.7$\times10^{2}$ THz), an FWHM of 2.1 nm (3.2 THz) with an initial power of P$_{L}$ = 100 mW and a beam diameter of 4 mm. However, P$_{L}$ is reduced to roughly P$_{o}$ = 25.9 mW inside to the ZS generating a intensity I$_{o}$ = 2.1$\times10^{2}$ mW/cm$^{2}$. For the experiment, we took into account two different dipole transitions of the Gd. The first transition is located at 443.063 nm and corresponds to the transition between the ground state with total angular momentum \textbf{J} = 2 and the excited state with \textbf{J} = 3, both with electronic spin number \textbf{S} = 4. These states have spectroscopic terms $^{9}$D$^{0}$ and $^{9}$D, respectively~\cite{NIST_ASD}. On the other hand, the second transition is located at $\lambda$ = 451.965 nm and corresponds to an intermediate transition between a state with J = 3 and a state with J = 2, with $^{9}$D$^{0}$~\cite{NIST_ASD}. The two transitions have measured lifetimes of $\tau_{443.06}$ = 13.7 ns and $\tau_{451.96}$ = 10.8 ns~\cite{Den_Hartog_2011_Gadolinium_Tau}; and saturation intensity of I$_{443.06}$ = 1.0$\times10^{2}$ mW/cm$^{2}$ and I$_{451.96}$ = 1.3$\times10^{2}$ mW/cm$^{2}$. This intensity values generate ratios of I$_{443.06}$/I$_{sat}$ = 1.5 and\break I$_{451.96}$/I$_{sat}$ = 2.7, respectively. The Doppler-broadening suffered due to the oven temperature for the 443.06 nm transitions was 15.4 GHz, and 15.2 GHz for the 451.96 nm transition. Finally, the detuning between the laser and the transition frequencies are $\Delta\nu_{443.06}$ = 6.0 THz and $\Delta\nu_{451.96}$ = 7.2 THz. The detuning between the laser and the transitions was adjusted by the Zeeman effect that generates the apparatus. Therefore, we propitiated the interaction with the 451.9 nm transition during the experiment, and used the 443.06 nm transition to prove our method for other electronic transitions and compare the results.

After the ZS section, we measured the momentum exchange of the Gd atoms with a QCM that has a surface detection area of approximately 5.0$\times10^{-1}$ cm$^{2}$. The QCM was coupled to a thickness monitor model Maxtek TM-200 and a frequency meter model Hewlett Packard 53131A universal counter. The QCM is a device normally used in thin-film growth processes, and its operating principle is based on monitoring the change of natural frequency oscillation values of a quartz crystal, which are modified due to mass deposition. This system allowed us to measure any frequency variation for the QCM with a resolution of $\pm$1 mHz. Finally, this experiment used a modification of the procedure presented by M. Guevara-Bertsch et. al~\cite{3_guevara2018energy} to measure the momentum exchange in the deposition process.    
\section{Use of the QCM to measure the deceleration of Gd atoms generated for a Zeeman slower}
\label{sec:Experimental_method}

\begin{figure}
\begin{center}
\includegraphics[width=1.0\textwidth]{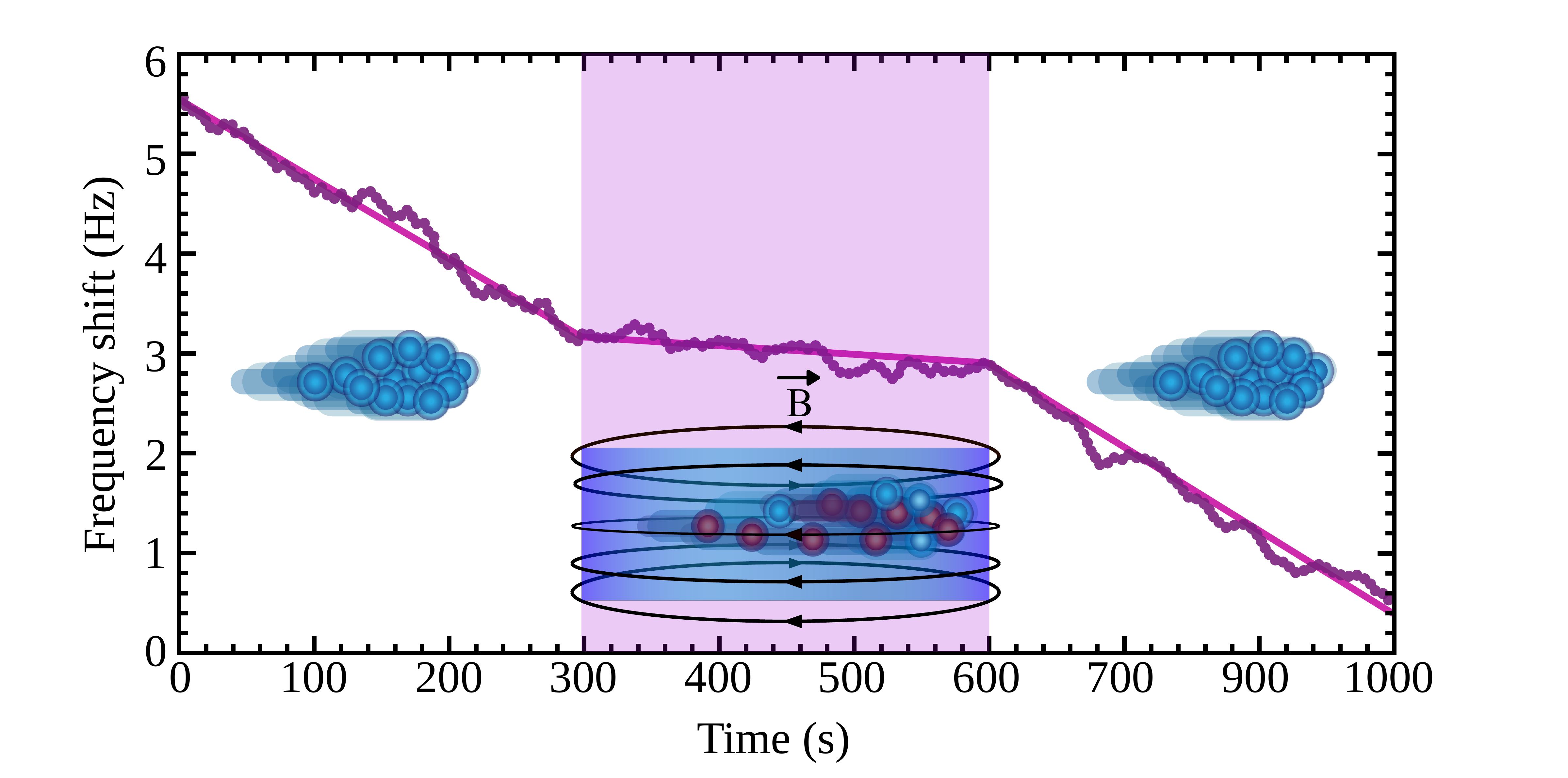}
\caption{Time evolution of the drop in the oscillation frequency of the QCM (purple dots). We create three stages of near five minutes each. In the white areas, we have the periods when our apparatus is off, and there no disturbances affecting the QCM frequency signal. In the shaded purple area, a perturbation of the frequency signal is introduced by the action of our Zeeman slower (ZS). The change in the slopes is more evident with the linear fits done (violet line). Moreover, in the shape, the laser beam an the magnetic field of the ZS generates the deceleration of the atoms. However, not all atoms are decelerated in the process, so the QCM frequency continues to drop but at a slower rate.}
\label{fig:Ajuste_Linear}
\end{center}
\end{figure}

To begin the experiment, we focus our attention on the measurement of the resonance frequency of the QCM. This frequency is perturbed only by two factors: the number of particles hitting per second and the heat transfered from the oven to the QCM. The first factor generates a momentum exchange that we want to measure with the QCM, while the second is a source of undesirable noise in the measurement that is necessary to reduce.

Therefore, in preparation for the experiment, we carry out heating of our vacuum setup from room temperature to $\sim$ 130 $^{0}$C, including annealing of the oven applying a current of 1 A for more than eight hours. This heating reduces the impurities making the system reach a base pressure of $10^{-9}$ Torr when the oven is off and increasing the mean free path of the atoms inside it. This heating allowed us to consider a continuous flow of atoms between the oven and the QCM that undergoes without significant change. Thus, it is possible to assume that the Gd atoms move out from the oven to the QCM without interaction with other particles in their trajectory, and only the desired effects can be considered. These results let us ignore any difference between the number of atoms that are emitted by our source and the number that reaches the QCM. In consequence, any resonance frequency perturbation of the QCM was created mostly by the arrival of the atoms to the surface of the crystal.
 
Similarly, from the point of view of the QCM, the direct exposure to the heat radiated by the oven, and the Joule effect of the ZS can increase the measured frequency values. Consequently, a shielding effect could be created that makes it impossible to determine the frequency variation due to the effects of our interest. This effect was reduced by slowly increasing the electric power up about 1000 W in the oven. Therefore, the temperature of the oven was increased at a rate that was necessary to produce a constant flux of atoms without creating an excessive and abrupt increase in the frequency of the QCM. Additionally, a continuous flow of water was used to extract the excess of heat from the QCM holder, and to stabilize the device's temperature in a short time and to cool the ZS coils. The flow of water not only reduced the temperature of the QCM but also reduces the noise of the frequency signal to an average calculated value of (45 $\pm$ 6) mHz during the experiment.

Once the optimal condition for the experiment was reached, the oscillation frequency of the QCM was continuously measured with a rate of 1 measurement per second. Hence, we established three measurement intervals of 5 minites each to generate enough experimental data to reduce the effect of any remaining unwanted disturbances. Frequency curves as shown in Fig.~\ref{fig:Ajuste_Linear}, were generated to characterize the frequency variation produced by the momentum exchange and deposition of the evaporated Gd atoms on the QCM.   

Consequently, we established a sequence of events that we need to carry out during the experiment excecution. This sequence consists of the following steps:

\begin{enumerate}
    \item First, we regulate the current and voltage of the source of atoms up to 1.4 A and 750 V, to start the evaporation of the sample of a Gd. At the same time, a flow of water is used to cold the ZS and the QCM, while a real-time plot was initiated to show the behavior of the frequency signal.
    
    \item Next, we wait around 10 minutes for the thermal stabilization of the QCM where we first see an increase in the frequency values due to the increase of the heat radiated by the oven. Then, the frequency signal begins to fall in an increasingly appreciable manner due to the hitting and deposition of the atoms in the QCM surface.

    \item As the first stage of measurement, we measure the fall in the frequency values produced by the incoming flux of Gd atoms in the QCM for 5 minutes with the ZS and the laser off. An experimental curve like the one shown in the left white zone of Fig.~\ref{fig:Ajuste_Linear} is generated and its slope ($s_{1}$) is taken for comparison with subsequent measurements.  

    \item While the flux of atoms continues to the QCM, we turn on the ZS and the laser at the same time and maintain these conditions for 5 minutes. We keep the laser intensity at a stable value of 91.2 mW and monitor that the current supply of the ZS coils is maintained. If any slowdown process was generated, a variation in the rate of incoming atoms to the QCM is produced. Therefore, a new slope ($s_{2}$) in the frequency curve is obtained, coinciding with what was observed in the central shaded region of Fig.~\ref{fig:Ajuste_Linear}.
    
    \item We turn off the ZS and the laser at the same time and wait for approximately 5 minutes in order to see a recovery in the falling of the frequency signal. We observe a downward slope ($s_{3}$) similar to the one in the right white section of Fig.~\ref{fig:Ajuste_Linear} with a trend close to that observed for the curve in step (iii).
	
	\item We repeat the cycle for different values of current applied to the ZS are attained.
	    
\end{enumerate}

The time of steps (iii), (iv), and (v) was adjusted $\sim$ 5 minutes to obtain enough data for the statistical analysis of each test. Finally, we ran tests where our ZS and laser were not turned on to rule out any electronic frequency disturbance from the QCM. This methodology was used to carry out a total number of 160 tests with the purpose of estimating the average perturbations in the falling slope of frequency curves produced by our apparatus for different current values in the ZS coils. 

\section{Characterization of the momentum exchange produced by Zeeman slower}
\label{sec:Variation_of_the_QCM_frequency_with_respect_Zeeman_Slower_effect}

As mentioned in previous studies~\cite{3_guevara2018energy}, it is possible to relate the measurements of the frequency of the QCM with the change in the momentum of the atoms. Therefore, in our experiment, the atoms exchange momentum with the QCM when deposited on its surface. The process causes a constant rate of fall of natural oscillation frequency value of the QCM that only depends on the number of atoms arriving at the QCM surface per unit of time. Consequently, if the power of the oven is constant, the number of atoms arriving and interacting with the QCM will be approximately constant. If this condition holds, the rate of fall of the frequency measured by the QCM will be nearly constant too. Thus, any perturbation observed over the rate of fall of the frequency values will be produced by the action of the laser and the Zeeman slower (ZS) over the atoms. 

Figure \ref{fig:Ajuste_Linear} shows the effect on the slope in the three steps of our experimental cycle. The steps 3 and 5 correspond to the white areas representing a free change of the frequency of the QCM. In these white areas, there is no deceleration of the atoms, and they remain unperturbed during the trajectory from the oven to the QCM. Instead, the shaded area that corresponds with step 4 where the slowdown process affect the atoms and generate a positive increase in the slope value of the frequency. As a result, the estimated variation of the momentum exchange between the atoms and the QCM in each experimental cycle is found. We carry out this estimation for each current value to find the maximum efficiency value of our ZS.

Consequently, we classified the data generated during each experimental cycle according to the current value used in the ZS coils. Then, each data set was separated in the three steps mentioned, and performed a linear fit on each one to obtain the value of $s_{1}$, $s_{2}$, and $s_{3}$. Hence, we established a criterion based on our observations made for the variation of the slopes which took two conditions for a successful test where $s_{1st}$ $<$ $s_{2}$ and $s_{3}$ $\sim$ $s_{1}$. Then, the percentage of variation between $s_{2}$ and $s_{1}$ ($\Delta s_{21}$ ), and between $s_{3}$ and $s_{1}$ ($\Delta s_{31}$) were estimated for each experimental cycle. As a result, when the laser was getting closer to resonance with the electronic transition due to the presence of a stronger Zeeman effect, an increase of $\Delta s_{21}$ was obtained as expected. This tendency continued until a maximum efficiency value for $\Delta s_{21}$ was reached. Finally, the laser came out of resonance because of the excessive Zeeman effect and $\Delta s_{21}$ falls again. On the other hand, we only expected a little variation of the values of $\Delta s_{31}$ because during step 3 and step 5 there is not a slowdown process; therefore, $s_{1}$ and $s_{3}$ must have similar values. 

To perform the analysis for the large number of tests carried out with different current values, a computer program was developed using the Python programming language. The program uses PyWavelets~\cite{7_lee2006pywavelets}, Pandas, Numpy~\cite{8_nelli2015python}, and Scipy~\cite{9_virtanen2020scipy} libraries, for cleaning and data processing of each experimental test. The program operated in the following way, the signal noise and the signal to noise ratio were reduced to average values of (26.7 $\pm$ 3.2) mHz and 55.3 $\pm$ 10.0, respectively. Second, each data set was separated and $s_{1}$, $s_{2}$, and $s_{3}$ was calculated using a linear fit. Finally, the program compared $s_{2}$ against $s_{1}$. The analysis was performed by matching $s_{1}$ to the 100$\%$ like undisturbed fall of frequency and also representing a maximum momentum exchange between the atoms and the QCM. Following, $s_{2}$ was calculated and compared against $s_{1}$ to obtain the variation percentage between both quantities. The result was a measurement of the reduction of the momentum exchange between the Gd atoms and the QCM, which was generated by the action of the magnetic field produced by the ZS during the cooling process. The program can be accessed at $https://github.com/Rocketman5990/Zeeman\_Project.git$, and it is open to download, use, modify and improve for free.

\begin{figure}
\begin{center}
\includegraphics[width=1.0\textwidth]{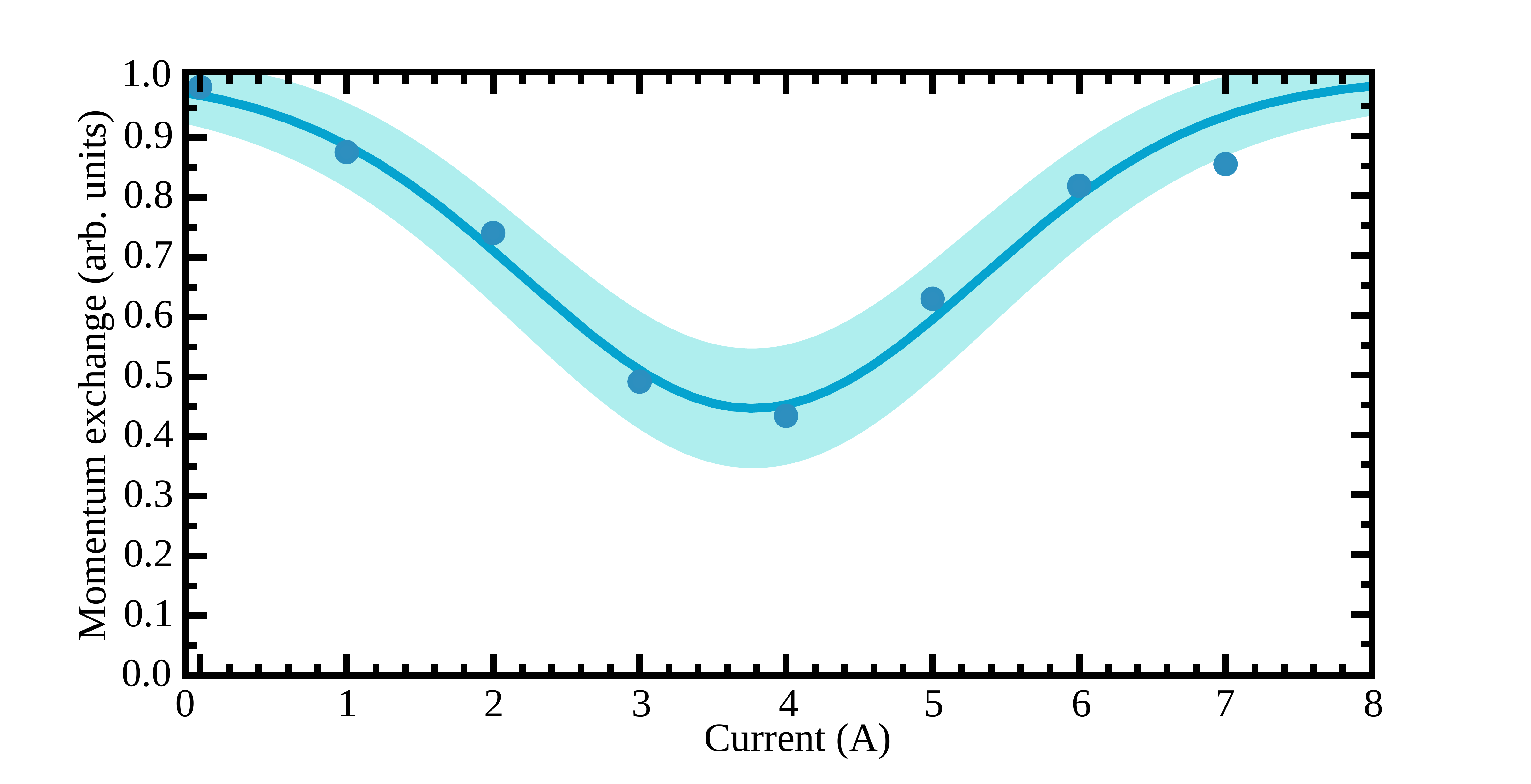}
\caption{Reduction percentages of momentum exchange concerning the applied currents values. The experimental points (blue dots) were obtained by calculating the percentage variation ($\Delta s_{21}$ ) between the slopes of the frequency curves obtained when the ZS was off ($s_{1}$) and when it was on ($s_{2}$), respectively.  The average percent obtained is related to the efficiency percentage when the ZS is operating with a given value of current. A Gaussian fit (green line) was used to obtain an approximate maximum efficiency value of (43.5 $\pm$ 6.4)$\%$, corresponding to an applied current of \break$\sim$ 3.8 A. The shaded area represents two standard deviations calculated for the Gaussian fit in the range of currents presented for each experimental point.}
\label{fig:Ajuste_Gaussiano}
\end{center}
\end{figure}

Figure \ref{fig:Ajuste_Gaussiano} presents the resulting efficiency curve for the experiment beginning with a current value of 0 A where no magnetic field was produced by the ZS. Here, there are an average change of (1.4 $\pm$ 0.49) $\%$ in $\Delta s_{21}$, which reflects that there is a minimum change in the momentum exchange between the atoms and the QCM. Then, the current of the ZS is increases, $\Delta s_{21}$ begins to increase, obtaining curves similar to Fig.~\ref{fig:Ajuste_Linear} with a more marked variation of $s_{1}$ and $s_{2}$. Thus, the percentage of momentum exchange between the Gd atoms and the QCM decreases concerning the increase in current due to the greather cooling process generated by an increasingly strong magnetic field. The behavior continues until the maximum efficiency current of 3.8 A is reached, producing a reduction of the\break (43.5 $\pm$ 6.4)$\%$ of the momentum exchange. As a consequence, the rate of atoms hitting the QCM per second is reduced to the minimum, and the variation $\Delta s_{21}$ reaches its maximum value. This result is shown in Fig.~\ref{fig:Ajuste_Gaussiano} and corresponds with the maximum efficiency of the slowdown process produced by our ZS. This value was affected by actual configuration of the apparatus, that is mentioned in section~\ref{sec:Experimental_setup} and might increase by testing other system configurations in future experiments.

Next, once the maximum efficiency current value is exceeded, the efficiency of our ZS began to decrease as expected. As a consequence, the momentum exchange increased and $\Delta s_{21}$ is reduced until a limit of detection for the QCM was reached. In this limit, although there could be changes in the value of the momentum exchanged between the QCM and the atoms, the magnitude of the natural noise of the QCM and $\Delta s_{21}$ begins to be comparable. For our experiment, the natural noise of the QCM was 45.1 mHz and began to be meaningful after 7 A. Any measurement after this point was not possible with the actual configuration of our experimental setup.

\section{Determination of the final velocity of Gd atoms and lifetimes for 443.06 nm and 451.96 nm transitions from the mechanic perturbation of the QCM frequency }
\label{sec:Determination of the final velocity of Gd(I) atoms and lifetime}

Due to the numerical method used, we took into account a Maxwell-Boltzmann velocity distribution (MBD) to describe the behavior of the atoms in the beam~\cite{foot,fox2006quantum}. This method allowed the study of the evolution of the MBD and his average velocity value through the Jaymes-Cuming Model (JCM) and the Fokker-Planck equation (FPE)~\cite{zoll}, mentioned in section~\ref{sec:Numerical_Model}. Consequently, we developed two different approaches to analyze the results obtained to demonstrated the cooling level of the Gd atoms achieved during the experiment. The first approach was based on the comparison between the final velocity found by our numerical model and the reduction percent of the atom velocity found by experimental data of the QCM; while the second was based on the estimation of the lifetime $\tau$ of the electronic transitions based on the experimental data obtained. In both cases, we used two different values of $\tau$ measured using time-resolved laser-induced fluorescence (TRLIF) on a slow atomic beam and reported by E. A. Den Hartog et al. in previous research~\cite{Den_Hartog_2011_Gadolinium_Tau}. The values reported for the used transition are $\tau_{443.06}$ = 13.7 ns for the 443.06 nm transition and $\tau_{451.96}$ = 10.8 ns for the 451.96 nm transition. These values were taken as a reference for our numerical model and used to compare our experimental values of $\tau$ ($\tau_ {e}$). 

On the first approach, the Gd atoms have an approximate initial experimental velocity of\break $\overline{v_{ie}}$ = 751.0 m/s when they leave out the oven at a temperature T $\backsim$ 4052 K, according to the MBD. On these initial conditions, the 443.06 nm transition is taken into account with a theoretical transition linewidth $\Gamma_{443.06}$ = 1/$\tau_{443.06}$ = 0.073 GHz. These considerations means that when the ZS worked at maximum efficiency with the tilted laser configuration, the atoms reached a maximum theoretical deceleration of $a_{443.06}$ = $\hbar k \Gamma_{443.06} / 2m_{Gd}$ = 3.33$\times10^{4}$ m/s$^{2}$. Then, the numerical model was used to study how the evolution of the MBD is affected by $a_{443.06}$ during the cooling process, as shown in Fig.~\ref{fig:fpe}. The result was a final approximated model velocity $\overline{v_{fm}}$ = 530 m/s that corresponds with a final model temperature T$_{fm}$ $\backsim$ 2084 K reached by the Gd in the experiment. The resulting value of $\overline{v_{fm}}$ indicates a reduction of 29.5 $\%$ in the velocity of the atoms according to our numerical model. At the same time, the velocity reduction percent obtained by the QCM data was 43.4 $\%$ taking into account $\overline{v_{ie}}$, indicating a final average experimental velocity\break $\overline{v_{fe}}$ = (326.7 $\pm$ 26.0) m/s. The final experimental temperature obtained for this velocity was T$_{fe}$ = (767.5 $\pm$ 123.7) K. The latter-results represent a difference of 32.5 $\%$ between the reduction value obtained by our numerical model and the experimental value obtained by the QCM. The percentages reduction indicated that a cooling process is being carried out in the Gd atoms beam using the $^{9}$D$^{0}_{2}$ - $^{9}$D$_{3}$ electronic transition. Additionally, with the 451,96 nm transition, for the use of the first approach was considered theoretical transition linewidth $\Gamma_{451.96}$ = 1/$\tau_{451.96}$ = 0.0926 MHz. During the cooling process, $\Gamma_{451.96}$ produces a deceleration $a_{451.96}$ = 4.1$\times10^{4}$ m/s$^{2}$. This deceleration generates a $\overline{v_{fm}}$ = 457.8 m/s according to the study of the evolution of the MBD, corresponding to a T$_{fm}$ = 1556.6 K. Therefore, for the transition of 451.96 nm, we obtain a reduction of 39.1 $\%$ of the velocity of the atoms according to the numerical model. This reduction percentage presents a difference of 10.0 $\%$ between the results obtained by our numerical model and the results obtained by the data of the QCM.

On the other hand, the second approach was based on the estimation of the value for $\tau_{e}$ from the experimental data obtained. The method considers the value of $\overline{v_{ie}}$ and the reduction of 43.4 $\%$ measured by the QCM for the velocity of the atoms to estimate $\overline{v_{fe}}$. Moreover, on this second approach, we stablished the condition $\overline{v_{im}}$ = $\overline{v_{ie}}$ and $\overline{v_{fm}}$ = $\overline{v_{fe}}$ as initial and final average velocity respectively for our numerical model. Thus, an estimated of $\tau_{e}$ = 8.2 ns was obtained using the maximum efficiency of the ZS measured by the QCM of 43.4$\%$ for the electronic transition used for the Gd atoms in the experiment. Then, it was carried out a comparison between $\tau_{e}$, and the reference values $\tau_{443.06}$ and $\tau_{451.96}$. These comparisons indicate a difference of 40.2$\%$ and 24.1$\%$, respectively between our experimental values and the reference values. Consequently, our approach indicates a higher probability that laser cooling occurs in the 451.9 nm transition, instead of the 443.06 nm transition, as expected. Despite the percentage differences obtained between $\tau_{e}$, $\tau_{443.06}$ and $\tau_{451.96}$, we consider that $\tau_{e}$ is consistent with the values reported by E. A. Den Hartog~\cite{Den_Hartog_2011_Gadolinium_Tau}. However, future improvements in our experiment could reduce the difference between the values obtained in our research and the values reported, increasing the sensitivity of the QCM for a better estimate of $\tau_{e}$. The QCM is normally used in solid-state physics, in processes of growing thin films, its frequency resolution is enough to measure mass variation in the range of $\mu$g; besides, it is possible use a QCM to measure the momentum exchange of a beam of atoms as mentioned in previous research~\cite{3_guevara2018energy}. Our results demonstrate that the evolution of the velocity distribution of a beam of atoms in laser cooling experiments might be determined by the implementation of the use of measurement of the momentum exchange using the QCM. Nevertheless, in our experiment, the 451 nm transition interacts only 15$\%$ with the laser emission curve according to our numerical model, and this affects the results obtained by lowering the efficiency of our experiment. Another aspect, is the distance between the QCM and the oven, near 1 m, which causes the beam of atoms to disperse; and therefore, not all the atoms that reach the QCM have interacted with the laser. In addition, the position of the QCM inside the vacuum system, as shown in Fig.~\ref{fig:Zeeman_working}, causes the laser to be tilted 1.3°, wich becomes another factor that affects our measurements. We think that the above factors make the results obtained in our experiment optimal only for the configuration used in our system, and more modifications are needed in order to achieve more accuracy for our measurements in future experiments and increase the sensibility. Nonetheless, the results showed that it may be possible to implement the use of the QCM with sufficient resolution to determine the speed change of a beam of atoms in the current configuration of our system.

However, our technique can be considered a novel way to measure the deceleration produced by the cooling process in a ZS and the lifetime of the electronic transitions based on the mechanical operating principles of the QCM. We compare our technique with TRLIF that can be carried out by two methods: pulse method and phase modulation-method~\cite{TRLIF_terzic2008development}. TRLIF has the ability to identify spectrally and temporally the different frequencies and lifetime of the electronic transitions of the atoms~\cite{TRLIF_plancque2003europium}. Hence, that the TRLIF is a predominant measure precedure in laser cooling experiments, but it may need different equipment like photo-multipliers and charge-coupled device (CCD). An example, other instruments used execute TRLIF are mentioned in others research~\cite{TRLIF_terzic2008development,TRLIF_yarlagadda2011radiative,TRLIF_plancque2003europium}, as part of different setups for independent purposes. On the other hand, the use of the QCM requires simple equipment that is more accessible for many not specialized laboratories. QCM is a well-known plug and play instrument that is normally used in experimental solid-state physics applications. Its use is simple and easy to implement for the measurement of the atoms' velocity changes by the measure of the momentum exchange. However, the QCM cannot be used to determine the excited atomic population that can be determined by the usual techniques like Doppler spectroscopy, saturation spectroscopy, two-photon spectroscopy, and optogalvanic spectroscopy~\cite{sushkov2013application,maguire2006theoretical,foot, sargsyan2018hyperfine}. Nonetheless, we think that the ability to measure the momentum exchange by using the QCM in a laser cooling experiment allow us to obtain enough data to characterize the deceleration effect generated in laser cooling experiments. Also, as our second analysis method showed, our technique could make it available to measure the lifetime of different electronic transition by a non-optical technique as a first approach, due to the remarked diferences in the reduction percent obtained for the 447.6 nm and the 451.96 nm transition. Thus, the QCM can be used to map the laser frequency range used to induce the electronic transitions in the laser cooling experiments for a certain type of atom. It could be possible to obtain the excitation frequencies for specific electronic transitions, in which a certain laser with a suitable frequency generates a deceleration that reduces the momentum exchange between the QCM and the atoms. These frequencies could coincide with the transition frequencies useful in laser cooling for the type of atom used. As a result, by improving the sensitivity level of the experimental system, the QCM might be used as a mapping sensor to the determination of the stronger electronic transition frequencies and their lifetimes. Thus, we consider the possible improvement of the QCM system for future exploration in the borderland field between solid-state physics and ultracold atoms physics for more novel applications in future projects. 

\section{Conclusions} 
In conclusion, we demonstrated the implementation of the QCM as an energy transfer sensor can detect the change in the velocity of a beam of atoms produced by a Zeeman slower (ZS) in laser cooling experiments. By the use of the technique used in previous research~\cite{3_guevara2018energy}, our experiment is based on the continuous monitoring of the perturbation of the oscillation of the frequency of the QCM. We also improve the analysis of the measurements of the frequency signal of the QCM by the use of post-processing software and an estimation of the change in the time dynamics between the hitting atoms and the QCM. Based on these measurements, we determined the maximum efficiency working parameters of our ZS for the actual configuration of our system. Moreover, we developed a numerical model to study the particle dynamics of a beam with a Maxwell-Boltzmann velocity distribution (MBD) during a laser cooling experiment. Thus, from this model, we obtained a theoretical prediction of the final average velocity for MBD of a Gd atoms beam that interacts with a laser beam of 447.10 nm and a magnetic field of a ZS. Furthermore, with the careful comparison between our numerical model and the experimental data, we also estimated the lifetime values of the 443.06 nm and 451.96 nm transition of the Gd. We compared our results with the values reported in previous research~\cite{Den_Hartog_2011_Gadolinium_Tau} obtaining 40.2$\%$ and 29.5$\%$ for the 443.06 nm and 451.96 nm transitions respectively. In consequence, the feasibility of the QCM as a mechanical sensor to measure first approximation quantum properties in laser cooling experiments is determined.

With this set of results, we propose the introduction of the QCM in laser cooling experiments as a simple and easy way to implement a plug and play apparatus study the change in particle dynamics during the cooling process as a first approximation. The QCM is a well-known device used in solid-state physics, and its main advantage is its easy operation. Additionally, the QCM is a good sensor for the study of the kinetic evolution of a distribution of particles, due to its mechanical operating principle that allow the measure of the momentum exchange. Therefore, the implementation of the QCM in the laser cooling experiments should contribute to developing a new applications of experimental methods between solid-state and ultracold atom physics. Our results confirm that the application of the QCM can make available an innovative mechanical method to study the kinetic evolution of the atoms inside a ZS, to obtain a deeper understanding of the cooling process. 

\section{Acknowledgments}  

The research group would like to thank to Milena Guevara-Bertsch for her work in the design and construction of our Zeeman slower during on the early stages of our experiment. We also want to thank Alejandro God\'{i}nez-Sand\'{i} for his participation in the development of the experimental procedures used in this investigation and his help in the initial steps of the execution of the experiment. Moreover, the authors are very grateful for the support given by the Vicerrector\'{i}a de Investigaci\'{o}n de la Universidad de Costa Rica to carry out this research work.


\bibliography{apssamp}

\end{document}